\documentclass[11pt]{amsart}
\usepackage{graphicx,amssymb,amsmath,amsthm}
\usepackage{enumerate}
\usepackage{comment,cite,color}
\usepackage{cite,color}
\usepackage{algorithmic}
\usepackage{mathrsfs}
\usepackage[ruled]{algorithm}
\usepackage{epsfig}
\usepackage{enumitem}

\newcommand{\PP}{\mathbb{P}}
\newcommand{\HH}{\mathbf{H}}
\newcommand{\innerp}[1]{\langle {#1} \rangle}

\newcommand{\abs}[1]{\lvert#1\rvert}
\newcommand{\absinn}[1]{\vert\langle {#1} \rangle\rvert}
\newcommand{\argmin}[1]{\mathop{\rm argmin}\limits_{#1}}

\newcommand{\lra}{{\longrightarrow}}
\newcommand{\eproof}{\hfill\rule{2.2mm}{3.0mm}}

\newcommand{\Proof}{\noindent {\bf Proof.~~}}

\newcommand{\BS}{{\mathbb S}}

\newcommand{\R}{{\mathbb R}}

\newcommand{\C}{{\mathbb C}}

\renewcommand{\eqref}[1]{(\ref{#1})}
\newcommand{\inner}[1]{\langle #1 \rangle}
\newcommand{\shsp}{\hspace{1em}}
\newcommand{\mhsp}{\hspace{2em}}

\newcommand{\NN}{{\mathcal N}}

\newcommand{\rank}{{\rm rank}}
\newcommand{\supp}{{\rm supp}}

\renewcommand{\H}{{\mathbb H}}
\newcommand{\qHH}{\widetilde{\HH}}

\newcommand{\MM}{\mathbf M}

\newcommand{\F}{{\mathcal F}}

\newtheorem{prop}{Proposition}[section]

\newtheorem{defi}{Definition}[section]

\newtheorem{theo}[prop]{Theorem}

\date{}

\begin{document}
\bibliographystyle{plain}
\title{Phase Retrieval for Sparse Signals}

\author{Yang Wang}
\thanks{Yang Wang was supported in part by the
       National Science Foundation grant DMS-0813750, DMS-08135022 and DMS-1043034.
       Zhiqiang Xu was supported  by NSFC grant 11171336, 11331012, 11021101 and National Basic Research
Program of China (973 Program 2010CB832702).}
\address{Department of Mathematics  \\ Michigan State University\\
East Lansing, MI 48824, USA}
\email{ywang@math.msu.edu}

\author{Zhiqiang Xu}
\address{LSEC, Inst.~Comp.~Math., Academy of
Mathematics and System Science,  Chinese Academy of Sciences, Beijing, 100091, China}
\email{xuzq@lsec.cc.ac.cn}

\maketitle

\begin{abstract}
The aim of this paper is to build up the theoretical  framework for the recovery of  sparse signals from the magnitude
of the measurement. We first investigate the minimal number of measurements for the success of the recovery of
sparse signals without the phase information. We completely settle the minimality question for the real case
and give a lower bound for the complex case.
We then study the recovery performance of the
$\ell_1$ minimization for the sparse phase retrieval problem.
 In particular, we present the null space property  which, to our knowledge, is the first
 sufficient and necessary condition for the success of $\ell_1$ minimization for $k$-sparse phase retrievable.
\end{abstract}

\section{Introduction}
\setcounter{equation}{0}

The theory of compressive sensing has generated enormous interest in recent years. The goal of
compressive sensing is to recover a sparse signal from its linear measurements, where the
number of measurements is much smaller than the dimension of the signal, see e.g. \cite{tao,tao1,donoho,CanTao05}.
The aim of this paper is to study the problem of
compressive sensing without the phase information. In this problem the goal is to
recover a sparse signal from the magnitude of its linear samples.

Recovering a signal from the magnitude of its linear samples, commonly known
as {\em phase retrieval}
or {\em phaseless reconstruction}, has gained considerable attention in recent years \cite{BCE06, BBCE07, CSV12, CESV12}. It has
important application in  X-ray imaging, crystallography, electron microscopy,
coherence theory and other applications. In many applications the signals to be
reconstructed are sparse. Thus it is
natural to extend compressive sensing to the phase retrieval problem.

We first introduce the notation and briefly describe the mathematical background of
the problem. Let $\F=\{f_1, f_2, \dots, f_m\}$ be a set of vectors in $\H^d$
 where $\H$ is either $\R$ or $\C$. Assume that $ x\in\H^d$ such that
 $b_j = |\inner{x, f_j}|$. The phase retrieval problem asks whether we can reconstruct
 $x$ from $\{b_j\}_{j=1}^m$. Obviously, if $y =cx$ where $|c|=1$ then
 $|\inner{y,f_j}|=|\inner{x,f_j}|$. Thus the best phase retrieval can do is to
 reconstruct $x$ up to a unimodular constant.

Consider the equivalence relation $\sim$ on $\HH:=\H^d$: $x \sim y$
if and only if there is a constant
$c\in \H$ with $|c|=1$ such that $x=cy$. Let $\qHH :=\HH/\!\!\sim$. We shall use
$\tilde{x}$ to denote the equivalent class containing $x$. For a give $\F$
in $\HH$ define the map $\MM_\F: \qHH \lra \R_+^m$ by
\begin{equation}  \label{1.1}
      \MM_\F(\tilde{ x}) = [|\inner{\tilde{x},f_1}|^2, \dots, |\inner{\tilde{x},f_m}|^2]^\top.
\end{equation}
The phase retrieval problem asks whether a $\tilde{ x}\in\qHH$ is uniquely determined
by $\MM_\F(\tilde x)$, i.e. $\tilde x$ is recoverable from $\MM_\F(\tilde x)$.
 We say a set of vectors $\F$ has the {\em phase retrieval property},
or is {\em phase retrievable},
if $\MM_\F$ is injective on $\qHH = \H^d/\!\!\sim$.

It is known that in the real case $\H=\R$ the set $\F$ needs to have at least
$m \geq 2d-1$ vectors to have the phase retrieval property; furthermore a generic
set of $m \geq 2d-1$ elements in $\R^d$ will have the phase retrieval property,
c.f. Balan, Casazza and Edidin \cite{BCE06}.
In the complex case $\H=\C$ the same question remains open,
and is perhaps the most prominent open problem in phase retrieval. It is known
that $m \geq 4d-2$ generic vectors in $\C^d$ has the phase retrieval property \cite{BCE06}.
It is also shown that there exists a set $\F$ with $m=4d-4$ vectors having the phase
retrieval property, c.f. Bodmann  and Hammen \cite{bodmann}. The current conjecture is that
phase retrieval property in $\C^d$ can only hold when $m \geq 4d-4$, and furthermore
any $m \geq 4d-4$ generic vectors in $\C^d$ have the phase retrieval property.

The aforementioned results concern with the general phase retrieval problem in $\H^d$.
In many applications, however, the signal $x$ is often sparse with $\|x\|_0 = k \ll d$.

We use the standard notation $\H^d_k$ to denote the subset of $\H^d$ whose elements
$x$ have $\|x\|_0 \leq k$. Let $\qHH_k$ denote $\H^d_k/\!\!\sim$. A set $\F$ of vectors in
$\H^d$ is said to have the {\em $k$-sparse phase retrieval property}, or is
{\em $k$-sparse phase retrievable}, if any
$\tilde x \in \qHH_k$ is uniquely determined by $\MM_\F(\tilde x)$. In other words, the map
$\MM_\F$ is injective on $\qHH_k$.
One may naturally ask: {\em how many
vectors does $\F$ need to have so that $\F$ is $k$-sparse phase retrievable?}

The best current results on the $k$-sparse
phase retrieval property are proved by  Li and Voroninski \cite{LiV}, which state
that $k$-sparse phase retrieval property can be achieved by having $m \geq 4k$ and $m\geq 8k$
vectors for the real and complex case, respectively (see also \cite{Eldar}).

In  Section 2, we prove sharper results for a set of vectors $\F$ to have
the $k$-sparse phase retrieval property.
In the real case $\H=\R$ we obtain a sharp result.
We show that for any $k<d$ the set $\F$ must have at least $m\geq 2k$
elements to be $k$-sparse phase retrievable. Furthermore, any
$m \geq 2k$ generic vectors will be $k$-sparse phase retrievable.
In the complex case $\H=\C$ we proved that
any $m \geq 4k-2$ generic vectors have the $k$-sparse phase retrieval property.
We conjecture that this bound is also sharp, namely for $k<d$
a set $\F$ in $\C^d$ needs at least
$4k-2$ vectors to have the $k$-sparse phase retrieval property.

A foundation of compressive sensing is built on the fact that the recovery of a sparse signal
from a system of under-determined linear equations is equivalent to finding the extremal value
of $\ell_1$ minimization under certain conditions. The $\ell_1$ minimization  is  extended
to the phase retrieval in \cite{algorithm3} and one also develops many algorithms to  compute  it (see \cite{algorithm1,algorithm2}).
However, there have been few theoretical results
on the recovery  performance of $\ell_1$ minimization for  sparse phase retrieval.
In Section 3,   we  present  the null space property, which, to our knowledge, is the first sufficient and necessary condition
for the success of $\ell_1$ minimization for $k$-sparse phase retrievable.
 If we take $k=d$, the null space property is reduced to  a  condition
of the frame $\F$ under which $\MM_\F$ is injective on $\C^d/\!\!\sim $ and we present it in Section 4.

\section{Minimal Sample Number for $k$-Sparse Phase Retrieval}
\setcounter{equation}{0}

In this section we study the problem of minimal number of samples (measurements)
required for $k$-sparse phase retrieval.
We shall introduce more notation here.  Often it is convenient to identify
a set of vectors $\F =\{f_1, f_2,\dots, f_m\}$ with the matrix $F=[f_1,f_2, \dots,f_m]$
whose columns are the vectors $f_j$. When $\F$ is a frame this is known as the
{\em frame matrix} of $\F$. We shall use the term frame matrix for $F$ regardless whether
$\F$ is a frame or not. Also for integers $n \leq m$ we use the notation $[n:m]$
to denote the set $\{n, n+1, \dots, m\}$. For $x\in \H^d$, we set $\abs{x}:=[\abs{x_1},\ldots,\abs{x_d}]$.
Similar with before, we let
$$
   \R^d_k\,:=\, \{x\in \R^d: \|x\|_0\leq k\}.
$$
Our first theorem completely settles the minimality
question for $k$-sparse phase retrieval in the real case $\H=\R$.

\begin{theo}  \label{theo-2.1}
Let $\F=\{f_1,\ldots,f_m\}$ be a set of vectors in $\R^d$. Assume that $\F$ is $k$-sparse
phase retrievable on $\R^d$. Then $m \geq \min\,\{2k, 2d-1\}$.
Furthermore, a set $\F$ of $m \geq \min\,\{2k, 2d-1\}$ generically chosen
vectors in $\R^d$ is $k$-sparse
phase retrievable.
\end{theo}

\Proof Note that the full sparsity case $k=d$ is already known: $m \geq 2d-1$ vectors
are needed for phase retrieval and a generic set of $\F$ with $m \geq 2d-1$ vectors
will have the phase retrieval property. So we will focus only on $k<d$.

We first prove that $m \geq 2k$. Assume $\F$ has $m<2k$ elements. We prove $\F$
does not have the $k$-sparse phase retrieval property
by constructing $x,y \in \R^d_k$ with
$|\inner{x,f_j}|=|\inner{y, f_j}|$ but $x\neq \pm y$.

We divide $\F$ into two groups: $\F_1 = \{f_j:~j\in[1:k]\}$ and
$\F_2 = \{f_j:~j\in[k+1:m]\}$. Let the corresponding frame matrices be
$F_1$ and $F_2$, respectively. Consider the subspace
$$
 W=\{[x_1,x_2, \dots, x_{k+1}, 0,\dots, 0]^\top\in \R^d:~x_1, \dots, x_{k+1}\in\R\}.
$$
For the first group $\F_1$, there exists a $u\in W\setminus \{0\}$ such that
$F_1^\top u=0$, i.e. $\innerp{f_j,u}=0$ for all $1\leq j\leq k$. This is because
$\dim(W)=k+1$ and there are only $k$ equations. Note also that there
are at most $k-1$ vectors in the second group $\F_2$ since $m-k<2k-k=k$.
Thus the solution space
$$
\{v\in W:~ F_2^\top v=0\}
$$
has dimension at least $2$. Hence, there exist linearly independent $\alpha, \beta\in W$
so that for all $t,s\in \R$
$$
     v=t\alpha+s\beta
$$
satisfies
$$
     F_2^\top v=0, \shsp\mbox{i.e.}\shsp \innerp{f_j,v}=0 \shsp\mbox{for}~j\in [k+1:m].
$$

Write $u=[u_1, u_2, \dots, u_d]^\top$ (where $u_j=0$ for $j>k+1$).
Since $\alpha$ and $\beta$ are linearly independent, we may without loss of generality assume
$[\alpha_1,\alpha_2]^\top $ and $[\beta_1,\beta_2]^\top$ are linearly independent, where
$\alpha=[\alpha_1,\ldots,\alpha_d]^\top$ and $\beta=[\beta_1,\ldots,\beta_d]^\top$. We first consider the
case where either  $u_1\neq 0$ or $u_2\neq 0$.
Then there exist $s_0, t_0\in \R$ with $(s_0,t_0)\neq (0,0)$ so that
\begin{eqnarray*}
   u_1 &=& t_0\alpha_1+s_0\beta_1,\\
  -u_2 &=& t_0\alpha_2+s_0\beta_2.
\end{eqnarray*}
Now set $\bar v = t_0\alpha + s_0\beta$ and
$$
    x:=u+\bar{v},\mhsp  y:=u-\bar{v}.
$$
Clearly $x, y \in \R^d_k$ since $\supp(x) \subseteq \{1,3, \dots, k+1\}$
and $\supp(y) \subseteq \{2,3, \dots, k+1\}$. Moreover
$$
   \innerp{f_j,x} = \innerp{f_j,u} + \innerp{f_j,\bar v} =
   \left\{\begin{array}{cl} \inner{f_j,u} &\shsp j\leq k\\
   \inner{f_j,\bar v} &\shsp j> k, \end{array}\right.
$$
and similarly
$$
   \innerp{f_j,y} = \innerp{f_j,u} - \innerp{f_j,\bar v} =
   \left\{\begin{array}{cl} \inner{f_j,u} &\shsp j\leq k\\
   -\inner{f_j,\bar v} &\shsp j> k. \end{array}\right.
$$
It follows that $|\innerp{f_j,x}|=|\innerp{f_j,y}|$ for all $j$ but $x\neq \pm y$. We next consider the
case where $u_1=u_2=0$.
Then there exist $s_0, t_0\in \R$ with $(s_0,t_0)\neq (0,0)$ so that
\begin{eqnarray*}
   0 &=& t_0\alpha_1+s_0\beta_1,\\
  1  &=& t_0\alpha_2+s_0\beta_2.
\end{eqnarray*}
Similar with before, we  set $\bar v = t_0\alpha + s_0\beta$ and
$$
    x:=u+\bar{v},\mhsp  y:=u-\bar{v}.
$$
Then $x,y\in \R_k^d$ and $|\innerp{f_j,x}|=|\innerp{f_j,y}|$ for all $j$ but $x\neq \pm y$.
Thus $\F$ does not have the $k$-sparse phase retrieval property in $\R_k^d$.

We next prove that a set $\F$ of $m \geq 2k$ generic vectors will
have the $k$-sparse phase retrieval property. Let us first fix $I,J\subset [1:N]$
 with $\#I=\#J=k$. The goal is to prove that
if  $x, y\in \R_k^N$ with ${\rm supp}(x)\subset I$ and ${\rm supp}(y)\subset J$ satisfying
\begin{equation}\label{2.1}
   \abs{\innerp{f_j,x}}^2=\abs{\innerp{f_j,y}}^2,\mhsp j=1,\ldots,m,
\end{equation}
then $x = \pm y$.
Equation (\ref{2.1}) implies that for all $j$ we have
\begin{equation}  \label{2.2}
  \innerp{f_j,x-y}\cdot \innerp{f_j,x+y}=0.
\end{equation}
Thus either $\inner{f_j,x-y}=0$ or $\inner{f_j,x+y}=0$.
Without loss of generality, we assume that
\begin{equation}  \label{2.3}
  \begin{aligned}
    \innerp{f_j,x-y} &= 0,  \shsp & j\in [1:n]\\
    \innerp{f_j,x+y} &= 0,  & j \in [n+1:m].
  \end{aligned}
\end{equation}
Set
$$
   L:=I\cap J\mhsp \mbox{and}\mhsp \ell:=\# L.
$$
For convenience we write
\begin{eqnarray*}
   x&=&u_x+v_x, \mhsp {\rm supp}(u_x)\subset L,\,\, {\rm supp}(v_x)\subset I\setminus L, \\
   y&=&u_y+v_y, \mhsp {\rm supp}(u_y)\subset L,\,\, {\rm supp}(v_y)\subset J\setminus L .
\end{eqnarray*}
We abuse the notation a little by viewing $v_x\in \R^{k-\ell}$ since it is supported
on $I\setminus L$ with $\# (I\setminus L)=k-\ell$.
Similarly we view $v_y\in \R^{k-\ell}$ and $ u_x, u_y\in \R^\ell$. Set
$$
   w_{-}:=u_x-u_y,\,\, w_+:=u_x+u_y,\,\, {\rm and}\,\,
   z:=\left[\begin{array}{c} v_x\\ v_y \\ w_-\\ w_+ \end{array} \right]\in \R^{2k}.
$$
Using the notions above, we have
\begin{equation} \label{2.4}
   \begin{aligned}
    \innerp{f_j,x-y} &=\innerp{f_j,v_x}-\innerp{f_j,v_y}+\innerp{f_j,w_-},\\
    \innerp{f_j,x+y}&=\innerp{f_j,v_x}+\innerp{f_j,v_y}+\innerp{f_j,w_+}.
   \end{aligned}
\end{equation}
Set $A:=F^\top$ where $F$ is the frame matrix of $\F$.
Combining (\ref{2.3}) and  (\ref{2.4}) now yields
\begin{equation}\label{2.5}
\left[
\begin{array}{cccc}
A_{[1:n], I\setminus L} & -A_{[1:n], J\setminus L} & A_{[1:n],  L} &0  \\
 A_{[n+1:m], I\setminus L}  &   A_{[n+1:m], J\setminus L}&0  &A_{[n+1:m],  L}
\end{array}
\right]
\left[
\begin{array}{c}
v_x\\ v_y \\ w_-\\ w_+
\end{array}
\right]=0,
\end{equation}
where for any index sets $J_1,J_2$ we use the notation $A_{J_1,J_2}$ to denote
the sub-matrix of $A$ with the rows indexed in $J_1$ and columns indexed in $J_2$.
To show $x = \pm y$ we only need to show that the linear equations (\ref{2.5})
force $v_x=0 ,v_y=0$ and either $w_-=0$ or $w_+=0$.

We first consider the case $n\geq 2k-\ell$. In this case, we consider only the
first set of equations (\ref{2.5})
\begin{equation}\label{2.6}
\left[
\begin{array}{ccc}
A_{[1:n], I\setminus L} & -A_{[1:n], J\setminus L} & A_{[1:n],  L}  \\
\end{array}
\right]
\left[
\begin{array}{c}
v_x\\ v_y \\ w_-
\end{array}
\right]=0.
\end{equation}
Note that the matrix
$$
\left[A_{[1:n], I\setminus L} ~ -A_{[1:n], J\setminus L} ~ A_{[1:n],  L}\right]
$$
has dimensions $n\times (2k-\ell)$. The elements are generically chosen. Thus it
has full rank $2k-\ell$. It follows that \eqref{2.6} has only trivial solution
$v_x=0, v_y=0$ and $w_-=0$. Hence $x=y$.

We next consider the case with $m-n\geq 2k-\ell$. Here we consider the second set
of equations (\ref{2.5}):
\begin{equation} \label{2.7}
\left[
\begin{array}{cccc}
 A_{[n+1:m], I\setminus L}  &   A_{[n+1:m], J\setminus L}  &A_{[n+1:m],  L}
\end{array}
\right]
\left[
\begin{array}{c}
v_x\\ v_y \\ w_+
\end{array}
\right]=0.
\end{equation}
The same argument used for the case $n\geq 2k-\ell$ now applies to yield
$v_x=0, v_y=0$ and $w_+=0$. Hence in this case $x=-y$.

We finally consider the case where $n<2k-\ell$ and $m-n<2k-\ell$.  In this case we must have
$$
   2k-\ell\,\, >\,\, m-n \,\,\geq\,\, 2k-n,
$$
and hence $n>\ell$. Similarly, we have $\ell<2k-n\leq m-n$. We argue that the rank of the matrix in
(\ref{2.5}) is $2k$ when $F^\top$ is generic. Let $B$ denote the matrix in
\eqref{2.5}. If $\rank(B) <2k$
then all $2k\times 2k$ sub-matrices of $B$
have determinant $0$. Note that each determinant is either identically $0$ or a nontrivial
polynomial of the entries of $F$. Hence if there exists a single example of a matrix $B$
with $\rank(B)=2k$ then $\rank(B)=2k$ for a generic choice of $F$.
We shall construct an example of such an $F$ with $\rank(B)=2k$. Set
\begin{equation*}
A_{[1:n],L}=\left[
\begin{array}{c}
 I_\ell\\
 0
\end{array}
\right],
A_{[n+1:m],L}=\left[
\begin{array}{c}
I_\ell\\ 0
\end{array}
\right],
\end{equation*}
\begin{equation*}
[A_{[1:n],I\setminus L}, -A_{[1:n],J\setminus L}]=\left[
\begin{array}{c}
0\\ H_1
\end{array}
\right],  [A_{[n+1:m],I\setminus L}, -A_{[n+1:m],J\setminus L}]=\left[
\begin{array}{c}
0\\ H_2
\end{array}
\right],
\end{equation*}
where $I_\ell$ denotes the $\ell\times\ell$ identity matrix.
With this choice, for almost all
$H_1\in \R^{(n-\ell)\times (2k-2\ell)},H_2\in \R^{(m-n-\ell) \times (2k-2\ell)}$
we have $\rank(B)=2k$.
The solution to (\ref{2.5}) is thus trivial, namely
$v_x=0$, $v_y=0$, $w_-=0$ and $w_+=0$. Thus $x=y=0$. The theorem is now proved.
\eproof

\vspace{3mm}

We next consider the complex case. Similar to the real case we set
$$
   \C^d_k\,:=\, \{x\in \C^d: \|x\|_0\leq k\}.
$$
  Then we have

\begin{theo}  \label{theo-2.2}
   A set $\F$ of $m \geq 4k-2$ generically chosen vectors in $\C^d$ is $k$-sparse
phase retrievable.
\end{theo}
\Proof  We shall identify $\F$ with $F$ where
$\F=\{f_1, f_2, \dots, f_m\}$ is the corresponding frame matrix,
$F=[f_{ij}]$. Following the technique in \cite{BCE06} we shall view $F$ as an element
in $\R^{2md}$. The goal here is to show that the set of matrices $F$ that are not
$k$-sparse phase retrievable has local real dimension strictly smaller than $2md$ provided
$m\geq 4k-2$.

For any subset of indices $I,J\subset [1:m]$ with $\#I=\#J=k$ let $G_{I,J}$
denote the set of matrices in $\C^{d\times m}$ with the following property:
There exist  $x,y\in \C^d$ where
${\rm supp}(x)\subset I$, ${\rm supp}(y)\subset J$ and
$x\neq cy$ with $|c|=1$ such that $\MM_\F(x) = \MM_\F(y)$, i.e.
$\abs{\innerp{f_j,x}}= \abs{\innerp{f_j,y}}$ for all $j$. Now if
$\MM_\F(x)=\MM_\F(y)$, the for any $a, \omega\in\C$ with $|\omega|=1$ we also
have $\MM_\F(ax) = \MM_\F(a\omega y)$. Thus for any $F\in G_{I,J}$ we may find
$x,y\in\C^d$ with $\MM_\F(x) = \MM_\F(y)$ such that
\begin{itemize}
\item ${\rm supp}(x)\subset I$, ${\rm supp}(y)\subset J$.
\item The first nonzero entry of $x$ is 1.
\item The first nonzero entry of $y$ is real and positive.
\end{itemize}

Let $X$ denote the subset of $\C^d$ consisting of elements $x\in\C^d$ whose
first nonzero entry is 1. Let $Y$ denote the subset of $\C^d$ consisting of elements
$y\in\C^d$ whose first nonzero entry, if it exists, is real and positive.
Note that in essence $X$ can be
viewed as the projective space $\PP^{d-1}\setminus \{0\}$ and
$Y$ can be viewed as the set $\C^d/\sim$.
Let $\C^d_I$ denote the set of vectors $x\in\C^d$ such that $\supp(x) \subseteq I$.
Now consider the set of $3$-tuples
$$
    {\mathcal A}_{I,J}\,\,:=\,\, \{(F,x,y)\}
$$
with the following properties:
\begin{itemize}
\item $x\in X\cap \C^d_I$ and $y\in Y\cap \C^d_J$.
\item $x\neq \omega y$ for any $\omega \in \C$ with $|\omega|=1$.
\item $\MM_\F(x) = \MM_\F(y)$.
\end{itemize}

Now the projection of ${\mathcal A}_{I,J}$ to the first component gives the
full set $G_{I,J}$. Each $(F,x,y) \in {\mathcal A}_{I,J}$ gives rise to the
constraints $|\innerp{f_j,x}| =|\innerp{f_j, y}|$ for $j \in [1:m]$, which
lead to the set of
quadratic equations in ${\rm Re}(f_{ij}), {\rm Im}(f_{ij})$ (by viewing $x$, $y$ as fixed)
\begin{equation}\label{eq:qua}
   \abs{\sum_{k=1}^Nf_{kj}x_k}^2=\abs{\sum_{k=1}^Nf_{kj}y_k}^2,\mhsp j=1,\ldots,m.
\end{equation}
Note that all equations are independent and each is non-trivial because $x\neq y$ in
$\C^d/\sim$. Thus for any fixed $x,y$ the
set of such $A=[f_{ij}]$ satisfying (\ref{eq:qua}) is a real algebraic variety of (real)
codimension $2md-m$. Hence, ${\mathcal A}_{I,J}$ has local dimension everywhere at most
\begin{eqnarray*}
& &2md-m+{\rm dim}_\R(X\cap \C^d_I)+{\rm dim}_\R(Y\cap\C^d_J)\\
& &=2md-m+2k-2+2k-1\\
& &=2md-(m-4k+3).
\end{eqnarray*}
It follows from $m\geq 4k-2$ that ${\mathcal A}_{I,J}$ has local (real) dimension at most $2md-1$.
Now $G_{I,J}$ is the projection of ${\mathcal A}_{I,J}$ onto the first component.
Thus, $G_{I,J}$ has dimension at most $2md-1$. In other words,
a generic $F\in \C^{d\times m}$ is not in $G_{I,J}$.

Finally, the set of $F\in \C^{d\times m}$ not having the $k$-sparse phase retrieval property
for $\C_k^d$ is the union of all $G_{I,J}$ with
$\#I=\#J=k$. It is a finite union. The theorem is now proved.
\eproof

\vspace{3mm}
\noindent
{\bf Remark.}~~Although the above theorem shows that in the complex case any
$m \geq 4k-2$ generically chosen vectors are $k$-sparse phase retrievable, it is unknown whether
$4k-2$ is in fact the minimal number required.  We conjecture that the minimal number of vectors needed for being
$k$-sparse phase retrievable is indeed $4k-2$. Note that it is obvious that the conjecture
holds for $k=1$.

\section{Null Space Property for Sparse Phase Retrieval}
\setcounter{equation}{0}

In this section, we investigate the performance of $\ell_1$ minimization for sparse phase retrieval with extending  the null space property in compressed sensing to the phase retrieval setting.
We first introduce the null space property in compressed sensing, and then extend it to the phase retrieval
setting on $\R_k^d$ and $\C_k^d$, respectively.

\vspace{2mm}

\subsection{Null space property}
A key concept in compressive sensing is the so-called {\em null space property} of a
matrix.
For a given frame $\F = \{f_1, \dots, f_m\}
\subset \H^d$, we use $F$ to denote the frame matrix.
 Let $\NN(F)$ denote the kernel of $F^\top$, i.e.,
 $$
 \NN(F)=\{\eta:\innerp{f_j,\eta}=0, j=1,\ldots,m\}.
 $$
 To state conveniently, when $F=\emptyset$, we set  $\NN(F):=\H^d$.
\begin{defi}
The matrix $F$ satisfies the {\em null space property of order $k$} if for any
nonzero $\eta=[\eta_1, \dots, \eta_d]^\top\in \NN(F)$ and any $T\subset [1:d]$ with $\# T \leq k$
it holds that
$$
     \|\eta_T\|_1 < \|\eta_{T^c}\|_1,
$$
where $T^c$ is the complementary index set of $T$ and $\eta_T$ is the restriction of $\eta$ to $T$.
\end{defi}
A fundamental result in compressed sensing is that a signal $x\in \H^d_k$ can be
 recovered via
the $\ell_1$-norm minimization if and only if the sensing matrix $A$ has the
null space property of order $k$ . We state it as follows (see \cite{null1,null2,null3,null4,null5}):

\begin{theo}\label{th:csnull}
Let $\F$ be a set of vectors in $\H^d$ and $F$ be the associated frame matrix.
Then $F$ satisfies the null space property of order $k$ if and only if
it has
$$
    \argmin{x\in \H^d}\bigl\{\|x\|_1:~F^\top x=F^\top x_0\bigr\}\,\,=\,\, x_0
$$
for  every $x_0\in \H_k^d$.
\end{theo}

\medskip

\subsection{The null space property for the real sparse phase retrieval }
Our goal here is to extend Theorem \ref{th:csnull} to the phase retrieval for the real signal.
For a given frame $\F = \{f_1, \dots, f_m\}$ and a subset $S$ of
$[1:m]$ we shall use $\F_S$ to denote
the set $\F_S:=\{f_j:~j\in S\}$. Similarly for the frame matrix we shall use
$F_S$ to denote the corresponding frame matrix of $\F_S$, i.e. the matrix whose columns are
the vectors of $\F_S$. We first consider the real case.

\begin{theo}\label{th:null}
    Let $\F=\{ f_1, f_2, \dots, f_m\}$ be a set of vectors in $\R^d$ and $F$ be the
associated frame matrix.  The following properties are equivalent:
\begin{itemize}
\item[\rm (A)]~~ For any $x_0\in \R_k^d$ we have
\begin{equation} \label{3.3}
   \argmin{x\in \R^d}\bigl\{\|x\|_1:~|F^\top x| = |F^\top x_0|\bigr\} = \{\pm x_0\},
\end{equation}
where $|F^\top x|=[\abs{\innerp{f_1,x}},\ldots, \abs{\innerp{f_m,x}}]^\top$.
\item[\rm (B)] ~~ For every
$S\subseteq [1:m]$ with $\# S\leq k$, it holds
$$
     \|u+v\|_1\,\,<\,\,\|u-v\|_1
$$
for all nonzero $u\in  {\mathcal N}(F_S)$ and $v\in \NN(F_{S^c})$
satisfying $\|u+v\|_0\leq k$.
\end{itemize}
\end{theo}
\Proof
First we show (B) $\Rightarrow$ (A). Let $b=[b_1, \dots, b_m]^\top:=\abs{F^\top x_0}$ where $x_0\in\R^d_k$.
For a fixed $\epsilon \in \{1,-1\}^m$ set
$b_\epsilon := [\epsilon_1b_1, \dots, \epsilon_m b_m]^\top$.
We now consider the following minimization problem:
\begin{equation}\label{3.4}
     \min \|x\|_1 \mhsp \mbox{ s.t. } \mhsp F^\top x=b_\epsilon.
\end{equation}
The solution to (\ref{3.4}) is denoted as $x_\epsilon$. We claim that for any
$\epsilon \in \{1,-1\}^m$ we must have
$$
    \|x_\epsilon\|_1\geq \|x_0\|_1
$$
if $x_\epsilon$ exists (it may not exist), and the equality holds if and only if $x_\epsilon=\pm x_0$.

To prove the claim let $\epsilon^*\in \{1,-1\}^m$ such that
$ b_{\epsilon_*}= F^\top x_0.$ Note that property (B) implies the classical null space property of order $k$.
To see this, for any nonzero $\eta\in \NN(F)$ and $T \subseteq [1:d]$ with $\#T \leq k$, set
$u: = \eta$ and $v:= \eta_T-\eta_{T^c}$. Let $S=[1:m]$. Then $u\in \NN(F_S)$ and $v\in \NN(F_{S^c})$.
The hypothesis of (B) now implies
$$
    2\|\eta_T\|_1 = \| u+v\|_1 < \|u-v\|_1 = 2\|\eta_{T^c}\|_1.
$$
 Consequently
we must have $x_{\epsilon^*}=x_0$ by Theorem \ref{th:csnull}. Now for any $\epsilon \in \{-1,1\}^m \neq \pm \epsilon^*$, if $x_\epsilon$
doesn't exist then we have nothing to prove. Assume it does exist.
Set $S_* :=\{j:~\epsilon_j = \epsilon^*_j\}$. Then
$$
   \innerp{f_j, x_\epsilon} = \left\{\begin{array}{cl} \innerp{f_j, x_0} & \shsp j \in S_*, \\
   -\innerp{f_j, x_0} &\shsp  j \in S_*^c. \end{array}\right.
$$
Set $u: = x_0 -x_\epsilon$ and $v: = x_0 +x_\epsilon$.
Clearly $u \in \NN(F_{S_*})$ and $v \in \NN(F_{S_*^c})$. Furthermore
$u+v= 2x_0 \in \R^d_k$. By the hypothesis of (B) we must have
$$
       2\|x_0\|_1 = \|u+v\|_1 < \|u-v\|_1 = 2\|x_\epsilon\|_1.
$$
This proves (A).

Next we prove (A) $\Rightarrow$ (B).
Assume (B) is false, namely, there exist nonzero $u\in \NN(F_S)$ and $v\in\NN(F_{S^c})$
such that $\|u+v\|_1 \geq \|u-v\|_1$ and $u+v\in\R^d_k$. Now set
$$
x_0 := u+v\,\,\in\,\, \R_k^d.
$$
Clearly,
$$
  \absinn{f_j, x_0} = \absinn{f_j,u+v}=\absinn{f_j,u-v},\mhsp j=1,\ldots,m
$$
since either $\innerp{f_j,u}=0$ or  $\innerp{f_j,v}=0$.
In other words, $|F^\top x_0| = |F^\top(u-v)|$. Note that $u-v \neq -x_0$,  for otherwise we would have
$u=0$, a contradiction. It follows from the hypothesis of (A) that we must have
$$
    \|x_0\|_1 = \|u+v\|_1 < \|u-v\|_1.
$$
This is a contradiction.
\eproof

\medskip

\subsection{The null space property for the complex sparse phase retrieval}
We now consider the complex case $\H=\C$. Throughout this subsection, we say that
${\mathcal S}=\{S_1,\ldots,S_p\}$ is a partition of $[1:m]$ if
$$
    S_j\subset [1:m],\shsp \bigcup_{j=1}^p S_j=[1:m] \shsp  \mbox{and}
    \shsp S_j\cap S_\ell=\emptyset \shsp \mbox{for all $ j\neq \ell$}.
$$
To state conveniently, we set
$\BS:=\{c\in \C: \abs{c}=1\}$
and
$$
\BS^m:=\{(c_1,\ldots,c_m)\in \C^m: \abs{c_j}=1, j\in [1:m]\}.
$$
Then we have:

\begin{theo}\label{th:cnull}
   Let $\F=\{ f_1, f_2, \dots, f_m\}$ be a set of vectors in $\C^d$ and $F$ be the
associated frame matrix. The following properties are equivalent.
\begin{itemize}
\item[\rm (A)]~~ For any $x_0\in \C_k^d$ we have
\begin{equation} \label{eq:cnull33}
   \argmin{\tilde x\in \C^d/\sim}\bigl\{\|x\|_1:~|F^\top x| = |F^\top x_0|\bigr\} =\tilde{x}_0,
\end{equation}
where $|F^\top x|=[\abs{\innerp{f_1,x}},\ldots, \abs{\innerp{f_m,x}}]^\top$ and $\tilde{x}_0$ denotes
the equivalent class $\{c x_0: c\in \BS\}$ in $\C^d/\!\!\sim$ containing $x_0$.
\item[\rm (B)] ~~ Suppose that
 $S_1,\ldots, S_p$ is any partition  of $[1:m]$ and that $\eta_j\in \NN({F_{S_j}})\setminus \{0\}$ satisfy
 \begin{equation}\label{eq:spc}
     \frac{\eta_1-\eta_\ell}{c_1-c_\ell}\,\,=\,\, \frac{\eta_1-\eta_j}{c_1-c_j}\,\,\in\,\,
       \C_k^d\setminus \{0\} \shsp \mbox{for all} \shsp \ell, j\in [2:p],
 \end{equation}
for some pairwise distinct  $c_1,\ldots,c_p\in \BS$.
Then
$$
\|\eta_j-\eta_\ell\|_1\,\, <\,\, \|c_\ell\eta_j-c_j\eta_\ell\|_1,
$$
for all $j,\ell\in [1:p]$ with $j\neq \ell$.
\end{itemize}
\end{theo}
\Proof
We first show (B)$\Rightarrow$ (A).
Let $b=[b_1, \dots, b_m]^\top:=\abs{F^\top x_0}$ where $x_0\in\C^d_k$.
For a fixed $\epsilon \in \BS^m$ set
$b_\epsilon := [\epsilon_1b_1, \dots, \epsilon_m b_m]^\top$.
We now consider the following minimization problem:
\begin{equation}\label{c3.4}
     \min \|x\|_1 \mhsp \mbox{s.t. } \mhsp F^\top x=b_\epsilon.
\end{equation}
The solution to (\ref{c3.4}) is denoted as $x_\epsilon$. We claim that for any
$\epsilon \in \BS^m$ we must have
$$
    \|x_\epsilon\|_1\,\,\geq\,\, \|x_0\|_1
$$
if $x_\epsilon$ exists (it may not exist), and the equality holds if and only if $\tilde{x}_\epsilon= \tilde{x}_0$.

To prove the claim let $\epsilon^*\in \BS^m$ such that
$ b_{\epsilon^*}= F^\top x_0.$
A similar argument  as the proof of Theorem \ref{th:null} shows
that property (B) implies the classical null space property of order $k$.
 Consequently
we must have $\tilde x_{\epsilon^*}=\tilde x_0$ by Theorem \ref{th:csnull}. Now we consider an arbitrary
$\epsilon \in \BS^m$.
If $\tilde \epsilon=\tilde{\epsilon^*}$, then $\tilde x_\epsilon=\tilde{x}_0$. So, we only consider the case where
$\tilde\epsilon\neq  \tilde{\epsilon^*}$. If $x_\epsilon$
does not exist then we have nothing to prove. Assume it does exist.
Set $c_j':=\epsilon_j/ \epsilon^*_j$
and $\eta_j':=c_j' x_{\epsilon^*}-x_{\epsilon}$ for $1\leq j \leq m$.
We can use $c_j'$ to define an equivalence relation on
$[1:m]$, namely $j \sim \ell $ if $c_j'=c_\ell'$. This equivalence relation leads to a partition
${\mathcal S}=\{S_1,\ldots,S_p\}$ of $[1:m]$. Now we set $c_j := c_\ell'$ where $\ell \in S_j$.
Clearly all $c_j$, $1 \leq j \leq p$, are distinct and unimodular.

Now set $\eta_j := c_j x_{\epsilon^*}-x_{\epsilon}$. Then we have
$$
  \eta_j\in {\mathcal N}(F_{S_j})\setminus\{0\}, \text{ for all }  j\in [1:p]
$$
and
$$
  \frac{\eta_1-\eta_j}{c_1-c_j}\,\,=\,\,  \frac{\eta_1-\eta_\ell}{c_1-c_\ell}\,\,\in\,\, \C_k^d, \mhsp\mbox{for all $j, \ell \in [2:p]$}.
$$
By the hypothesis of (B) we must have
$$
       \abs{c_j-c_\ell}\cdot \|x_0\|_1 = \|\eta_j-\eta_\ell\|_1 <
       \|c_\ell \eta_j-c_j\eta_\ell\|_1 = \abs{c_j-c_\ell}\cdot \|x_\epsilon\|_1,
$$
which implies that
$$
   \|x_0\|_1\,\,<\,\,  \|x_\epsilon\|_1.
$$
This proves (A).

We next prove (A) $\Rightarrow $ (B). Assume (B) is false, namely,
there exist nonzero $\eta_j\in \NN(F_{S_j}), j\in [1:p]$ satisfying  (\ref{eq:spc})  but
$$
\|\eta_{j_0}-\eta_{\ell_0}\|_1\,\, \geq \,\, \|c_{\ell_0}\eta_{j_0}-c_{j_0}\eta_{\ell_0}\|_1
$$
for some distinct  $j_0,\ell_0\in [1:p]$.
Note that (\ref{eq:spc}) implies that
\begin{equation}\label{eq:jlmn}
\frac{\eta_j-\eta_\ell}{c_j-c_\ell} \,\,=\,\,\frac{\eta_m-\eta_n}{c_m-c_n}\,\,\in\,\, \C_k^d\setminus \{0\},
\end{equation}
for  all $j,\ell, m, n\in [1:p]$ with $j\neq \ell$ and $m\neq n$,
Without loss of generality, we assume that $j_0=1,\ell_0=2$, i.e.,
\begin{equation}\label{eq:cnull}
\|\eta_1-\eta_2\|_1\,\, \geq \,\, \|c_2\eta_1-c_1\eta_2\|_1.
\end{equation}
Set
$$
   x_0\,\,:=\,\,{\eta_1-\eta_2},
$$
and (\ref{eq:jlmn}) implies that $x_0\in \C_k^d\setminus\{0\}$.
We claim that
\begin{equation}\label{eq:claim}
   \abs{\innerp{f_j,x_0}}=\abs{\innerp{f_j,\eta_1-\eta_2}}=\abs{\innerp{f_j,c_2\eta_1-c_1\eta_2}}, \quad \text{ for all }j\in [1:p].
\end{equation}
Note that $x_0$ is $k$-sparse. Combining  (\ref{eq:claim}), (\ref{eq:cnull}) and (\ref{eq:cnull33}) now yields
$$
   cx_0=c\eta_1-c\eta_2=c_2\eta_1-c_1\eta_2
$$
for some  $c\in \BS$. Consequently we obtain
$$
   (c-c_2)\eta_1=(c-c_1)\eta_2,
$$
which implies that
\begin{equation}\label{eq:ceta21}
\eta_2=\frac{c-c_2}{c-c_1}\eta_1.
\end{equation}
Here,  note that $c\notin\{c_1,c_2\}$, for otherwise we will have either $\eta_1=0$ or $\eta_2=0$.
Combining (\ref{eq:spc}) and (\ref{eq:ceta21}) leads to
\begin{itemize}
  \item $\eta_1$ is $k$-sparse;
  \item  for all $j\in [2:p]$, $\eta_j$  and $\eta_1$ are linear dependent and hence $\eta_1\in \NN(F_{S_j})$.
\end{itemize}
And hence we have  $F^\top \eta_1=0$. By the hypothesis of (A) and $\eta_1\in \C_k^d$ we have
$\eta_1=0$. A contradiction.

We remain to prove (\ref{eq:claim}). First, when $j\in S_1\cup S_2$, (\ref{eq:claim}) holds, since either
$\innerp{f_j,\eta_1}=0$ or $\innerp{f_j,\eta_2}=0$.  We consider the case where $j\in S_3$. Set
$y_0:=\frac{\eta_1-\eta_2}{c_1-c_2}$. Then (\ref{eq:jlmn}) implies that
\begin{eqnarray*}
\frac{\eta_1-\eta_3}{c_1-c_3}=\frac{\eta_2-\eta_3}{c_2-c_3}=y_0
\end{eqnarray*}
and hence
\begin{eqnarray*}
\eta_1&=&(c_1-c_3)y_0+\eta_3,\\
\eta_2&=&(c_2-c_3)y_0+\eta_3.
\end{eqnarray*}
Note that $\innerp{f_j,\eta_3}=0$ with $j\in S_3$. Then
\begin{eqnarray*}
\abs{\innerp{f_j, c_2\eta_1-c_1\eta_2}}
&=&\abs{\innerp{f_j,c_2(c_1-c_3)y_0-c_1(c_2-c_3)y_0}}\\
&=&\abs{\innerp{f_j, c_3(c_1-c_2)y_0}}=\abs{\innerp{f_j,\eta_1-\eta_2}}=\abs{\innerp{f_j,x_0}}.
\end{eqnarray*}
Using a similar argument, we easily prove the claim for $j\in S_4,\ldots,S_p$.
\eproof

\vspace{3mm}
\noindent
{\bf Remark.}~~Theorem \ref{th:null} extends results for
the null space property of order $k$ in compressive sensing to phase retrieval. It will be very interesting
for constructing matrix $A\in \R^{m\times d}$ with $m\asymp k \log d$ satisfying {\rm (B)} in Theorem \ref{th:null}.

\section{Null space property for general phase retrieval}

Theorem \ref{th:null} and Theorem \ref{th:cnull} present the null space property for the phase retrievable on $\R_k^d$ and $\C_k^d$, respectively. In phase retrieval, one
is also interested in the condition under which $F$ is phase retrievable on $\R^d$ or $\C^d$. For the real case,
such a condition is presented in  \cite{BCE06}:
\begin{theo}\label{th:bce06}{\rm (\cite{BCE06})}   Let $\F=\{ f_1, f_2, \dots, f_m\}$ be a set of vectors in $\R^d$ and $F$ be the
associated frame matrix.
The following properties  are equivalent:
\begin{itemize}
\item[{\rm ({\rm A})}] $F$ is  phase retrievable on $\R^d$;
\item[{\rm ({\rm B})}] For every subset $S\subset\{1,\ldots,m\}$, either $\{f_j\}_{j\in S}$ spans $\R^d$ or  $\{f_j\}_{j\in S^c}$
spans $\R^d$.
\end{itemize}
\end{theo}
We next consider the complex case.
Motivated by  Theorem \ref{th:cnull}, we can present  the null space property under which $F$ is phase retrievable on $\C^d$. It can be considered
as an extension of Theorem \ref{th:bce06}:
\begin{theo}
    Let $\F=\{ f_1, f_2, \dots, f_m\}$ be a set of vectors in $\C^d$ and $F$ be the
associated frame matrix.   The following properties are equivalent:
\begin{itemize}
\item[\rm (A)]~ $F$ is   phase retrievable on $\C^d$;
\item[\rm (B)]
 Suppose that
 $S_1,\ldots, S_p$ is any partition  of $[1:m]$.
 There exists no  $\eta_j\in \NN({F_{S_j}})~\setminus~\{0\}, j=1,\ldots,p,$ such that
\begin{equation}\label{eq:spcc}
 \frac{\eta_1-\eta_\ell}{c_1-c_\ell}\,\,=\,\, \frac{\eta_1-\eta_j}{c_1-c_j}\,\, \neq \,\, 0 \,\,\text{ for all } \ell, j\in [2:p],
 \end{equation}
for some  pairwise distinct   $c_1,\ldots,c_p\in \BS$.
\end{itemize}
\end{theo}
\Proof
We first  prove (A) $\Rightarrow $ (B). Assume (B) is false, namely,
there exist nonzero $\eta_j\in \NN(F_{S_j}),\, j\in [1:p],$ satisfying  (\ref{eq:spcc}).
Set
$$
   x_0\,\,:=\,\,{\eta_1-\eta_2}.
$$
Using a similar method as the proof of (\ref{eq:claim}), we obtain that
$$
   \abs{\innerp{f_j,x_0}}=\abs{\innerp{f_j,\eta_1-\eta_2}}=\abs{\innerp{f_j,c_2\eta_1-c_1\eta_2}}, \text{ for all } j\in [1:p].
$$
Then, according to (A) and the definition of phase retrievable, we have
$$
   cx_0=c\eta_1-c\eta_2=c_2\eta_1-c_1\eta_2
$$
for some unimodular constant $c\in \BS\setminus\{c_1,c_2\}$,
which implies that
\begin{equation}\label{eq:ceta41}
\eta_2=\frac{c-c_2}{c-c_1}\eta_1.
\end{equation}
Combining (\ref{eq:spcc}) and (\ref{eq:ceta41}), we obtain that, for all $j\in [2:p]$, $\eta_j$  and $\eta_1$ are linear dependent and hence $\eta_1\in \NN(F_{S_j})$. So, $F^\top \eta_1=0$. The (A) implies that $\eta_1=0$, a contradiction.

We next show (B)$\Rightarrow$ (A).
Set $b=[b_1, \dots, b_m]^\top:=\abs{F^\top x_0}$ where $x_0\in\C^d\setminus \{0\}$.
For a fixed $\epsilon \in \BS^m$ set
$b_\epsilon := [\epsilon_1b_1, \dots, \epsilon_m b_m]^\top$.
We now consider the solution to
\begin{equation}\label{c4.4}
   \mhsp F^\top x=b_\epsilon.
\end{equation}
The solution to (\ref{c4.4}) is denoted as $x_\epsilon$. We claim that if  $x_\epsilon$ exists
then $\tilde{x}_\epsilon=\tilde{x}_0$, which implies (A).
Recall that $\tilde{x}_0$ denotes
the equivalent class $\{c x_0: c\in \BS\}$ in $\C^d/\!\!\sim$ containing $x_0$.
To prove the claim let $\epsilon^*\in \BS^m$ such that
$ b_{\epsilon^*}= F^\top x_0.$
The (B) implies that the rank of $F$ is $d$.
 Consequently
we must have $x_{\epsilon^*}= x_0$. Now we consider an arbitrary
$\epsilon \in \BS^m$.
If $\tilde \epsilon=\tilde{\epsilon^*}$, then $\tilde x_\epsilon=\tilde{x}_0$.
To this end, we only need prove that $x_\epsilon$ does not exist if
$\tilde\epsilon\neq  \tilde{\epsilon^*}$. Assume $x_\epsilon$ does exist.
Set $c_j':=\epsilon_j/ \epsilon^*_j$
and $\eta_j':=c_j' x_{\epsilon^*}-x_{\epsilon}$ for $1\leq j \leq m$.
We can use $c_j'$ to define an equivalence relation on
$[1:m]$, namely $j \sim \ell $ if $c_j'=c_\ell'$. This equivalence relation leads to a partition
${\mathcal S}=\{S_1,\ldots,S_p\}$ of $[1:m]$. Now we set $c_j := c_\ell'$ where $\ell \in S_j$.
Clearly all $c_j$, $1 \leq j \leq p$, are distinct and unimodular.
Now set $\eta_j := c_j x_{\epsilon^*}-x_{\epsilon}$. By definition for all $1\leq j \leq p$ we have
$$
  \eta_j\in {\mathcal N}(F_{S_j})\setminus \{0\}
$$
and
$$
  \frac{\eta_1-\eta_j}{c_1-c_j}=  \frac{\eta_1-\eta_\ell}{c_1-c_\ell}\neq 0 \mhsp\mbox{for all $j, \ell \in [2:p]$},
$$
which contradicts with (B). And hence $x_\epsilon$ does not exist if $\tilde\epsilon\neq \tilde\epsilon^*$.
This proves (A).

\eproof

\end{document}